\input phyzzx
\baselineskip 24pt plus 1pt minus 1pt
\hfill\vbox{\hbox{UUPHY/95/10}
\hbox {March, 1995}
\hbox {hep-th/9503216}}\break
\NPrefs
\let\refmark=\NPrefmark
\def\define#1#2\par{\def#1{\Ref#1{#2}\edef#1{\noexpand\refmark{#1}}}}
\def\con#1#2\noc{\let\?=\Ref\let\<=\refmark\let\Ref=\REFS
         \let\refmark=\undefined#1\let\Ref=\REFSCON#2
         \let\Ref=\?\let\refmark=\<\refsend}


\define\GUTH
A. H. Guth, Phys. Rev. {\bf D23} (1981) 347;
A. Albrecht and P. J. Steinhardt, Phys. Rev. Lett. {\bf 48}
(1982) 1220; 

\define\LINDE
For a review, see for example, A. D. Linde, "Particle Physics and
inflationary cosmology" (Harwood Academic Publishers, New York, 1990).

\define\CAMPBELL
B. A. Campbell, A. Linde and K. A. Olive, Nucl. Phys. {\bf B355}
(1991) 146; R. Brustein and P. J. Steinhardt, Phys. Lett. {\bf B
302} (1993) 196.

\define\AGUTH
A. H. Guth, Phys. Rev. {\bf D23} (1981) 347.

\define\ALINDE
A. D. Linde, Phys. Lett. {\bf B108} (1982) 389.

\define\ALBRECHT
A. Albrecht and P. J. Steinhardt, Phys. Rev. Lett. {\bf 48}
(1982) 1220.

\define\ADLINDE
A. D. Linde, Phys. Lett. {\bf B129} (1983) 177.

\define\DLA
D. La and P. J. Steinhardt, Phys. Rev. Lett. {\bf 62} (1989) 376.

\define\BRANS
C. Brans and R. H. Dicke, Phys. Rev. {\bf 24} (1961) 925.

\define\GREEN
M. B. Green, J. Schwartz and E. Witten, Superstring theory, vol. 1 and 2
(Cambridge Univ. Press, Cambridge, 1987).

\define\GASPERINI
B. A. Campbell, A. Linde and K. A. Olive, Nucl. Phys. {\bf B355} 
(1991) 146;
A. A. Tseytlin, Int. J. Mod. Phys. {\bf D1} (1992) 223;
A. A. Tseytlin and C. Vafa, Nucl. Phys. {\bf B372} (1992) 443;
M. Gasperini and G. Veneziano, Mod. Phys. Lett. {\bf A8} (1993) 3701;

\define\VENE
G. Veneziano, Phys. Lett. {\bf B265} (1991) 287;
A. A. Tseytlin, Mod. Phys. Lett. {\bf A6} (1991) 1721;
A. A. Tseytlin and C. Vafa, Nucl. Phys. {\bf B372} (1992) 443;
M. Gasperini and G. Veneziano, Astropart. phys. {\bf 1} (1993) 317.

\define\BRUSTEIN
R. Brustein and G. Veneziano, Phys. Lett. {\bf B329} (1994) 429.

\define\KALOPER
N. Kaloper, R. Madden and K. A. Olive, Nucl. Phys. {\bf B452}
(1995) 1995; \break 
Phys. Lett. {\bf B371} (1996) 34.

\define\GAN
C. G. Callan and Z. Gan, Nucl. Phys. {\bf B272} (1986) 647;
S. R. Das and B. Sathiapalan, Phys. Rev. Lett. {\bf 56} (1986) 2664;
A. A. Tseytlin, Phys. Lett. {\bf B264} (1991) 311.

\define\MAHAPATRA
V. Alan Kostelecky and M. J. Perry, Nucl. Phys. {\bf B414} ((1994) 174;
E. Raiten, Nucl. Phys. {\bf B416} (1994) 881; S. Mahapatra and S. 
Mukherji, Mod. Phys. Lett. {\bf A 10} (1995) 183.

\define\KS
V. A. Kostelecky and S. Samuel, Phys. Rev. {\bf D42} (1990) 1289.

\define\VENEZIANO
G. Veneziano, Phys. Lett. {\bf B265} (1991) 287.   

\title{\bf{A TACHYONIC EXTENSION OF THE STRINGY NO-GO THEOREM}}
\author{Swapna Mahapatra, Karmadeva Maharana and Lambodar 
P. Singh\foot{e-mail: swapna@iopb.ernet.in; kmu@utkal.ernet.in; lps@
utkal.ernet.in}} 
\address{Department of Physics, Utkal University, Bhubaneswar-751004,
India.}

\abstract
We investigate the tachyon-dilaton-metric system to study the 
"graceful exit" problem in string theoretic inflation, where tachyon
plays the role of the scalar field.  
>From the phase space analysis, we find that  
the inflationary phase does not smoothly
connect to a Friedmann-Robertson-Walker (FRW) expanding universe,
thereby providing a simple tachyonic extension of the recently proved 
stringy no-go theorem. 
\endpage
The expansion of the Universe including an inflationary phase
is considered to be an interesting and successful model in solving
various problems of standard cosmology \GUTH. The various variants of 
the inflationary scenarios are based on the existence of a generic 
scalar field \LINDE.  
A standard problem
in the inflationary cosmology is the "graceful exit" problem, which
relates to the fact that inflation never ends since the percolation 
of the true vacuum cannot exceed the expansion of the false vacuum
\CAMPBELL.
Analysis of standard problems in inflationary 
cosmology in a (super)string theory has led to interesting results. 
Superstring theory contains a 
scalar field known as dilaton in its spectrum, which is supposed to 
provide the necessary dynamics for inflation. A lot of work has been done
in past few years to investigate the role of dilaton in 
string inflationary cosmology \GASPERINI and the conclusion seems to be
that the presence 
of dilaton alone does not provide an adequate inflationary model.  
In recent times, use of duality symmetry in string cosmology 
has led to interesting results \VENE. Using this symmetry, Brustein 
and Veneziano have analyzed the "graceful exit" problem in the 
dilaton-gravity system in the string frame \BRUSTEIN. They have found 
that the transition 
from an accelerated inflationary phase to a standard FRW cosmological 
phase is not possible in the weak curvature regime. In two recent 
papers\KALOPER, the authors have reexamined the graceful exit problem
in string cosmology by including an axion field and a general axion-
dilaton potential. Their phase space analysis leads to a no-go
theorem, which rules out the possibility of a branch change necessary 
for graceful exit. In this paper, we
have considered the effect of tachyon field (which arises in the spectrum 
of bosonic string theory) on the string theoretic inflation. We have 
analysed the tachyon-dilaton-metric system to examine the 
graceful exit problem, that is to understand whether tachyon can play
a role in allowing the inflationary phase to end and connect smoothly 
to a Friedmann-Robertson-Walker (FRW) expanding universe. Our negative 
conclusion gives a tachyonic extension of the stringy no-go theorem 
established by Kaloper etal\KALOPER. 

The beta function analysis including the tachyon field has been 
considered before\GAN. It is known that tachyon destabilizes the 
canonical $26$ dimensional vacuum. But it is quite possible that 
the collective effects may stabilize the closed bosonic string 
in another ground state containing a nonzero tachyon expectation 
value. Upto cubic order in tachyon field, the string field theory 
gets a contribution to the tachyon potential of the form,
$$
V(\hat T) = - {2\over\alpha'} \hat T^2 + {\hat g\over 6} \hat T^3
\eqn\one$$
Here, $\hat T$ is the tachyon field appearing in string field theory
and $\hat g$ is the three point tachyon coupling at zero momentum. 

We start with the low energy string effective action involving the
metric $G_{\mu\nu}$, dilaton $\Phi$ and tachyon $T$. 
For the time dependent backgrounds, the equations of motion become 
quadratic so 
that solutions come in pairs corresponding to the two signs or 
branches and these two solutions are related through duality 
transformations. The positive branch solution describes accelerated 
expansion (contraction) corresponding to the pre big-bang phase and the 
negative branch solution describes deccelerated expansion (contraction)
corrsponding to the post big-bang phase of decreasing curvature. 
So the question is to see if the dynamics allows the two branches 
to join 
smoothly, so that at some later time, the universe
expands as a regular FRW universe. This picture has been analysed for
the dilaton driven inflation. It has been observed in ref.\BRUSTEIN\ that
though branch change is possible in this 
context, a second branch change also occurs 
which forbids  complete graceful exit from the accelerated inflation. 
This result was again confirmed by establishing an exact no-go 
theorem\KALOPER. From the phase space analysis, we find that the 
exact no-go
theorem is also valid in our case and hence tachyon 
can not induce a smooth branch change necessary for a 
successful graceful exit.

The four dimensional low energy string effective action is given by,
$$
S_{eff} = -{1\over 16\pi\alpha'}\,\int d^4 x {\sqrt G}e^{\Phi}\,
\lbrack - R - (\nabla\Phi)^2 + (\nabla T)^2 + 2 V(T) 
\rbrack\eqn\two
$$
The one equations of motion   
for the metric, tachyon and dilaton are 
respectively given by,
$$
R_{\mu\nu} = {\nabla}_{\mu}{\nabla}_{\nu}\Phi + 
{\nabla}_{\mu}T{\nabla}_{\nu}T\eqn\thre$$
$$
{\nabla}^2 T + {\nabla}_{\mu}\Phi {\nabla}^{\mu}T = V'(T)\eqn\four$$
$$ 
R - (\nabla\Phi)^2 - 2 {\nabla}^2\Phi - (\nabla T)^2 - 2 V(T) = 0
\eqn\five$$
where, $V'(T) = {\partial V(T)\over \partial T}$ and $V(T)$ is the tachyon 
potential. 
The cosmological evolution is described through solutions
of the dilaton-tachyon-gravity equations of motion \MAHAPATRA. We look 
for solutions of the equations of motion by starting with an ansatz for 
a flat, isotropic, FRW type of metric given by,
$$
d s^2 = - d t^2 + a^2(t) d x_i d x^i\eqn\seven
$$
where, $a(t)$ is the scale factor and $i = 1, 2, 3$. With this 
ansatz, the $tt$ and $ii$ components of the metric equation are 
respectively given by, 
$$
3 {\ddot a\over a} + \ddot\Phi + (\dot T)^2 = 0\eqn\eit
$$
$$
a \ddot a + 2 \dot a^2  + a \dot a \dot\Phi = 0\eqn\nine
$$
Tachyon and dilaton equations are respectively given by, 
$$
\ddot T + 3 {\dot a\over a} \dot T + \dot\Phi \dot T + V'(T) = 0
\eqn\ten$$
$$
(\dot\Phi)^2 + \ddot\Phi + 3{\dot a\over a} \dot\Phi - 2 V(T) = 0
\eqn\eleven$$

The extremum of the tachyon potential ($V'(T) =
0$) admits $T = T_0$ (constant) as a consistent solution apart from 
the case of $T = 0$\MAHAPATRA. We note that, for $T = T_0$, 
$V(T_0)\neq 0$
and $V''(T_0)$ is positive which ensures a stable solution. The Hubble 
parameter is defined as $H = {\dot a\over a}$.  
 
Now we rewrite the above equations in terms of Hubble parameter
$H$ and $\dot H$. After some algebraic manipulation, equations\eit,
\nine and \ten give the following two equations:
$$
\dot H = \pm H {\sqrt{3 H^2 + 2 V(T) + {\dot T}^2}}\eqn\twelve
$$
$$
\dot\Phi = - 3 H \mp {\sqrt{3 H^2 + 2 V(T) + {\dot T}^2}}\eqn\thirtn
$$
Note that these two equations are analogous to the previous 
axion-dilaton case (except for the derivative of the potential with respect 
to dilaton term and a term linear in $\rho$ in $\dot H$ equation as in 
our case, the derivative term in tachyon 
in the action does not have a dilaton factor in front and tachyon 
potential is independent of $\Phi$) by redefining our $\Phi = - 2\Phi$.
Finally, $T$ equation is given by,
$$
\ddot T \pm\lbrack{\sqrt{3 H^2 + 2 V(T) + {\dot T}^2}}\rbrack \dot T
+ V'(T) = 0\eqn\fourtn
$$

Now the solutions exhibit two different branches, called the positive
(+) branch and the negative (-) branch, according to the sign 
chosen (simultaneously for all the equations). We also get two more 
solutions due to the sign ambiguity 
of the initial value of $H$. The solutions obtained in this 
way, are related to each other by time reversal and scale factor 
duality (SFD) \VENEZIANO. The equations of motion remain invariant 
under these
symmetries. SFD transformations on various fields are given by, 
$$
\Phi \rightarrow \Phi + 6 \ln a;\qquad a \rightarrow {1\over a};
\qquad H \rightarrow - H;\qquad T \rightarrow T.\eqn\fiftn
$$

The solutions in the negative branch describes deccelerated expansion
($H > 0$, $\dot H < 0$) or deccelerated contraction ($H < 0$, $\dot 
H > 0$), 
depending on the initial sign of $H$. This branch can be 
joined smoothly to a standard radiation dominated FRW expanding 
universe and the solution is a stable one. On the other hand, the 
solution in the positive branch describes either accelerated expansion
($H > 0$, $\dot H > 0$) or accelerated contraction ($H < 0$, 
$\dot H < 0$), again depending on the initial sign 
of $H$. We see that since this branch accomodates only accelerated 
solutions, it does not join with the 
FRW expanding universe. A smooth joining in this case can happen only 
if the dynamics allows the positive branch to change over to the 
negative branch. 

>From the above analysis, we see that the equation governing a smooth 
transition is given by,
$$
3 H^2 + 2 V(T) + \dot T^2 = 0\eqn\eitn
$$
Since $H^2$ and $\dot T^2$ are positive, eqn.\eitn demands that 
$V(T)$ has to be 
negative for some value of $T$.  
Equation \eitn ensures continuity
in $\dot\Phi$. As we know that $\dot H$ remains invariant under 
the product of scale factor duality and time reversal transformation,
it is expected that solutions with same sign of $H$ can change to 
one another for maintaining continuity of $H$. Equation \eitn
describes a curve in phase space and for negative potential, this 
curve resembles an "Egg", which is assumed to be in the weak coupling
region. The egg function is defined as,
$$
e = {\sqrt{3 H^2 + 2 V(T) + {\dot T}^2}}\eqn\nintn$$
The egg is the region where $e\leq 0$. The condition for 
continuity in $\Phi$ (eqn. \thirtn) 
means that a branch change can occur only when $e = 0$.
In order to investigate the dynamics of the trajectories, we now consider
the four dimensional phase space of the problem which consists of 
$H$, $\Phi$, $T$ and $\dot T$. Taking a time derivative of the egg function 
expression and 
substituting the equations of motion, we obtain,
$$\pm \dot e = 3 H^2 + {\dot T}^2 = 2 H\dot\Phi - \dot H + 
{\dot T}^2\eqn\twntsevn$$
for the positive and negative branches respectively (we have already
redefined our $\Phi$ as $- 2\Phi$ in order to compare with the 
axion case). From the above equations, we can immediately see that for the 
positive 
branch, the derivative of the egg function is always non-negative 
(the expression after first equality) and hence there can be no branch 
change from the positive to the negative one. This gives a simple proof 
of the no-go theorem obtained by Kaloper etal. This can also be seen in 
the following way. As in the axion case, integrating equation 
\twntsevn w.r.t. time, 
the trajectory equation can be wriiten as, 
$$
\pm(e(t_2) - e(t_1)) + H(t_2) - H(t_1) = 2 \int_{t_1}^{t_2} H d\Phi
+ \int_{t_1}^{t_2} {\dot T}^2 dt\eqn\twnteit$$   

This is the same expression as the axion case (where ${\dot T}^2$
is replaced by the corresponding axion term $\rho$ without the multiplying 
dilaton factor), from which
a no-go theorem for branch change has been proved. 
Solving the equation for $\dot\Phi = 0$ (eqn.\thirtn ) in association
with the fact that $H$ is positive or negative for a point in 
the $H-\Phi$ plane to be above or below the egg respectively, 
the first term in eqn.\twnteit is positive definite as it represents
the area of the projection of the $\Phi$ trajectories in the 
$H-\Phi$ plane. 
The second term is positive by definition. Thus a branch change, which
requires l. h. s. to be negative, is ruled out much like the axion 
case \KALOPER. 
This implies that a positive trajectory entering the egg from above 
leaves as $(+)$ since conversion to negative trajectory is ruled out 
and a $(+)$ trajectory entering the egg from below leaves it as positive
since it can not hit the egg at all. 

Next, we examine the case when the positive trajectory hits the egg not above
or below but at the boundary where $H = 0$. We note that in this case,
$\pm\dot e = \rho$ (where $\rho = {\dot T}^2$). As the right hand side is 
positive by definition,
the positive trajectory is repelled away from the egg,
thus precluding any chance of a branch change. For 
$\rho = 0$, however this argument breaks down. The lowest nonvanishing 
time derivative of the egg function $e$ in this case is the triple 
derivative given by,
$$
\pm e^{(3)} = 2 \left({\partial V\over{\partial T}}\right)^2,\eqn\fifty$$
hence, $\pm e^{(3)} \geq 0$. Once again the positive trajectory is 
repelled from the egg forbiding a branch change. Towards an analysis of 
the case when the tachyon potential develops inflection points at the
boundary of the egg, it is observed that,
$$
H^{(n)} = 0, \qquad f^{(6)} = 80 {\left({\partial V\over
{\partial T}}\right)}^4\eqn\fif$$
where, $(n)$ stands for the $n$th derivative with respect to time 
and $f = 2\Lambda + \rho$. So for ${\partial V\over{\partial T}} = 0$,
the trajectory passing through $e = H = f = 0$ is a constant and does not 
lead to a branch change. From the above analysis, we note that the egg 
is incapable of effecting a branch change. Hence
tachyon can not induce a graceful exit from the inflationary 
phase, thereby providing a tachyonic extension of the previously extablished
no-go theorem.   

To summarize, we have analysed the tachyon-dilaton-gravity system
to understand the graceful exit problem in string cosmology. It has been 
shown in ref.\BRUSTEIN\ that a satisfactory answer to graceful 
exit problem can not be obtained in the dilaton-gravity system. 
This negative result has been confirmed in an axion-dilaton-metric
system by proving the no-go theorem for the branch change necessary
for a successful graceful exit \KALOPER.
We have closely followed the approach of Kaloper etal 
for investigating the dynamics of the trajectory and have extended the
the proof of the no-go theorem to the tachyonic case. 

\noindent{\bf Acknowledgement:}
We are thankful to the anonymous referees for many valuable suggestions 
and comments.  
S. M. would like to thank C. S. I. R. for the pool grant 
no.13(6764-A)/94.

\refout
\end